\documentclass[fleqn,10pt]{SelfArx}
\usepackage{lipsum}

\definecolor{color1}{RGB}{0,0,90} 
\definecolor{color2}{RGB}{0,20,20} 

\usepackage{hyperref} 
\hypersetup{hidelinks,colorlinks,breaklinks=true,urlcolor=color2,citecolor=color1,linkcolor=color1,bookmarksopen=false,pdftitle={Title},pdfauthor={Author}}

 \renewcommand{\vec}[1]{\mbox{\boldmath $#1$}}

 \def\gsim{\lower.4ex\hbox{$\;\buildrel >\over{\scriptstyle\sim}\;$}}
 \def\lsim{\lower.4ex\hbox{$\;\buildrel <\over{\scriptstyle\sim}\;$}}

 \def\adsr{AdSpR}
 
 \def\aap{A\&A}
 \def\apj{ApJ}
 \def\apjl{ApJL}
 
 \def\an{AN}
 \def\apss{Ap\&SS}
 \def\ga{Ge\&Ae}
 \def\gafd{GApFD}
 
 \def\mnras{MNRAS}
 \def\nat{Nature}

 \def\sci{Science}
 \def\sp{SoPh}
 
 \def\paj{AstL}

\JournalInfo{Astronomy Letters, 2017, Vol. 43, No. 4} 
\Archive{}
\PaperTitle{
A joined model for solar dynamo and differential rotation
}
\Authors{L.~L.~Kitchatinov\textsuperscript{1,2}* and A.~A.~Nepomnyashchikh\textsuperscript{1}}
\affiliation{\textsuperscript{1}\textit{Institute for Solar-Terrestrial Physics, Lermontov Str. 126A, Irkutsk, 664033, Russia}}
\affiliation{\textsuperscript{2}\textit{
Pulkovo Astronomical Observatory, Pulkovskoe Sh. 65, St. Petersburg, 196140, Russia
}}
\affiliation{*\textbf{E-mail}: kit@iszf.irk.ru}

\Keywords{dynamo - Sun: magnetic fields - Sun: rotation}
\newcommand{\keywordname}{Keywords}

\Abstract{
A model for the solar dynamo, consistent in global flow and numerical method employed with the differential rotation model, is developed. The magnetic turbulent diffusivity is expressed in terms of the entropy gradient, which is controlled by the model equations. The magnetic Prandtl number and latitudinal profile of the alpha-effect are specified by fitting the computed period of the activity cycle and the equatorial symmetry of magnetic fields to observations. Then, the instants of polar field reversals and time-latitude diagrams of the fields also come into agreement with observations. The poloidal field has a maximum amplitude of about 10 Gs in the polar regions. The toroidal field of several thousand Gauss concentrates near the base of the convection zone and is transported towards the equator by the meridional flow. The model predicts a value of about $10^{37}$~erg for the total magnetic energy of large-scale fields in the solar convection zone.
}

\begin{document}

\flushbottom 

\maketitle 


\thispagestyle{empty} 

\section{Introduction} 
Related theories of large-scale magnetic fields - the dynamo theory - and large-scale flows in the Sun and stars use similar methods but are largely uncoordinated. This refers to both the analysis of basic effects responsible for magnetic activity and global flows and to development of their quantitative models. The coordination of models for stellar dynamos and differential rotation is tempting for at least two reasons.

Solar dynamo models have advanced to close agreement with observations. Their extension to stars seems to be a natural next step. For the case of the solar dynamo, the differential rotation, and recently the meridional flow as well, can be prescribed after seismological inversions (Wilson et al. 1997; Schou et al. 1998; Rajaguru \& Antia 2015). This is hardly possible for stars, however. There are data on stellar surface rotation (e.g., Barnes et al. 2005; Croll et al. 2006; Walker et al. 2007) but distributions of global flows inside the convective envelopes is still beyond the reach of asteroseismology. The flows have to be defined with a theoretical model.

Such an approach also helps reduce uncertainties in specifying the dynamo parameters. This is not only because the differential rotation and meridional flow are now defined consistently. Our model of large-scale flows (Kitchatinov \& Ole\-mskoy 2011a, 2012a) does not follow the usual practice of prescribing the turbulent transport coefficients also, but defines them in terms of the entropy gradient (via superadiabaticity of stratification), which is one of the dependent variables of the model. The turbulent magnetic diffusion in the joined\footnote{We call the model  {\sl joined} but not {\sl unified} because unification implies taking into account the magnetic field influence on the flow, which is not the case with this paper.} model for dynamo and differential rotation proposed will be defined similarly. Two parameters of our model are, however, still uncertain (these are the ratio $\mathrm{Pm} = \nu_{_\mathrm{T}}/\eta_{_\mathrm{T}}$ of turbulent viscosity to magnetic diffusivity and the $n_\alpha$-pa\-ra\-me\-ter in the dependence of the $\alpha$-effect on latitude $\lambda$: $\alpha \sim \sin\lambda\ \cos^{n_\alpha}\lambda$). These two parameters are specified based on the condition that the simulated cycle period and equatorial symmetry of the magnetic field fit observations. It can be noted that the fitting brings other model characteristics - the instants of polar field reversals, the time-latitude diagram of the radial field, the ratio of polar to toroidal field amplitudes - in correspondence with observations.

The dynamo model of this paper is close to its former version (Kitchatinov \& Olemskoy 2012b). However, apart from its being matched with the differential rotation model, a significant difference from the former version is our allowance for the anistropy of turbulent transport coefficients induced by rotation. The former model, aimed at achieving though rough correspondence with observations, did not take into account anisotropy and other effects of secondary importance. Now, having achieved the aim, consideration of anisotropy, which might be consequential (Kitchatinov 2002), becomes justified.

The primary motivation for the model of this paper was our plan to apply the model to sun-like stars. Uncertain parameters of the model were specified by fitting observations of solar activity. The results of such \lq calibration' of the model seem to deserve this separate publication.
\section{Differential rotation}
Toroidal fields in stellar dynamos are generated by the differential rotation. Com\-pu\-ta\-ti\-ons of the differential rotation demand simultaneous computations of meridional flow and heat transport (Kitchatinov 2016). Contemporary models, therefore, define consistently the distributions of angular velocity, meridional circulation and specific entropy in the stellar convection zone.

The numerical model in the present paper is very close to that of Kitchatinov \& Olemskoy (2011a, 2012a). These publications and references therein provide a detailed description of the model which is not repeated here. The present paper, however, involves three modifications related to a slight elaboration of the model.

1. The equation for the meridional circulation,
\begin{eqnarray}
    \frac{\partial\omega}{\partial t} &+& r \sin\theta\ \mbox{\boldmath $\nabla$}\cdot
    \left(\mbox{\boldmath $V$}^\mathrm{m} \frac{\omega}{r\sin\theta}\right)\ +\
    {\cal D}\left({\vec V}^\mathrm{m}\right)
    \nonumber \\
    &=& \sin\theta\ r{\partial\Omega^2\over\partial z}\
    -\ {g\over c_{\rm p} r}{\partial S\over\partial\theta},
    \label{1}
\end{eqnarray}
is complemented by the term nonlinear in this circulation (the second term on the left-hand side of the equation). In this equation, the usual spherical coordinates ($r,\theta,\phi$) are used, ${\vec V}^\mathrm{m}$ is the meridional flow velocity, $\omega = (\mbox{\boldmath $\nabla$}\times{\vec V}^\mathrm{m})_\phi$ is the azimuthal vorticity, $\Omega$ is the angular velocity, $S$ is the specific entropy, $\partial/\partial z = \cos\theta\partial/\partial r - r^{-1}\sin\theta\partial/\partial\theta$ is the spatial derivative along the rotation axis, and ${\cal D}({\vec V}^\mathrm{m})$ accounts for the contribution of the turbulent viscosities   (viscous drag to the meridional flow). Turbulent viscosity in rotating fluid is anisotropic so that the explicit expression for ${\cal D}({\vec V}^\mathrm{m})$ is rather bulky (it can be found in the Appendix of the paper by Kitchatinov \& Olemskoy 2011a). This modification of the model is of minor significance. It leads to a change in the results of less than one percent and is involved in order to avoid any further discussions of a possible role of the now included nonlinearity and also for completeness of the model.

2. The heat transport equation,
\begin{equation}
    \rho T\left(\frac{\partial S}{\partial t} + {\vec V}^\mathrm{m}\cdot{\vec\nabla}S\right)
    = -{\vec\nabla}\cdot{\vec F} + R_{ij}\frac{\partial V_i}{\partial r_j},
    \label{2}
\end{equation}
now takes the thermal--kinetic energy exchange into account (the second term on the right-hand side). $\vec V$ is the velocity of the large-scale axisymmetric flow,
\begin{equation}
    {\vec V} = {\vec e}_\phi r\sin\theta\Omega(r,\theta)
    + \frac{1}{\rho} {\vec\nabla}\times\left( {\vec e}_\phi\frac{\psi(r,\theta)}{r\sin\theta}\right),
    \label{3}
\end{equation}
where ${\vec e}_\phi$ is the azimuthal unit vector, $\psi$ is the stream-function for the meridional flow, ${\vec F} = {\vec F}^\mathrm{rad} + {\vec F}^\mathrm{conv}$ is the heat flux consisting of the radiative and convective parts, $R_{ij} = -\rho\langle u_i u_j\rangle$ is the Reynolds stress tensor for the velocity $\vec u$ of turbulent convection, the angular brackets mean averaging, and repetition of subscripts implies summation. The last term in Eq.~(\ref{2}) guarantees conservation of total (kinetic plus thermal) energy and is included for model consistency. The Reynolds stress accounts, in particular, for the effect of turbulent viscosities. Easy estimations can show that the heating power for dissipation of the differential rotation by the eddy viscosity of $\nu_{_\mathrm{T}} \sim 10^{13}$~cm$^2$/s amounts to  several percent of solar luminosity. Our extension of Eq.~(\ref{2}) is, nevertheless, of minor consequence for the differential rotation model. The point here is that the Reynolds stress contains a non-dissipative part responsible for the differential rotation support (the $\Lambda$-effect; R\"udiger 1989) in line with the turbulent viscosities. In a steady state, the total - volume integrated - heat release by turbulent viscosities and heat sinks for support of the large-scale flow balance each other. The balance does not hold locally and the extension of Eq.~(\ref{2}) influences the results. As expected by Durney (2003), however, the influence is small (changes the results within 1\%).

\begin{figure}\centering
\includegraphics[width=7 truecm]{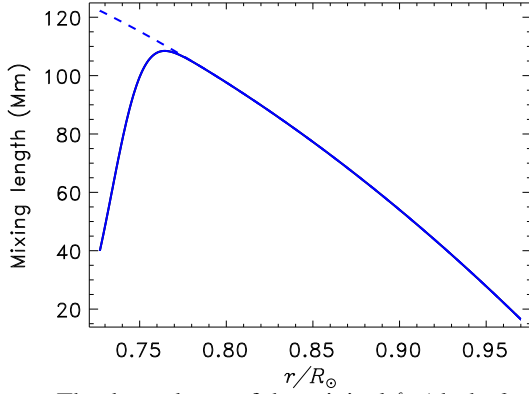}
\caption{The dependence of the original $\ell_0$ ({\sl dashed}) and cor\-rec\-ted
    $\ell$ (full line) mixing lengths on the heliocentric dis\-tance. The correction concerns the scale reduction near the base of the convection zone. The scale difference vani\-shes far from the base.
    }
    \label{f1}
\end{figure}

3. Our most significant modification of the differential rotation model is allowance for a reduction in the spatial scale of the turbulence near the base of the convection zone. The correlation length $\ell$ of turbulent flow (the mixing-length) is usually prescribed to be proportional to the pressure scale height $H_\mathrm{p} = - P/(\mathrm{d}P/\mathrm{d}r)$:
\begin{equation}
    \ell_0 = \alpha_\mathrm{MLT}H_\mathrm{p} .
    \label{4}
\end{equation}
The mixing-length (\ref{4}) increases with depth and exceeds 100 Mm near the base of the convection zone (Fig.\,1). It seems plausible, however, that the mixing-length should decrease on approaching the radiation zone where convection does not penetrate. In order to take into account the near-base scale reduction, we apply the equation
\begin{equation}
    \ell = \ell_\mathrm{min} + \frac{1}{2}\left(\ell_0 - \ell_\mathrm{min}\right)\bigg[
    1 + \mathrm{erf}\left(\left({r}/{R_\odot} - x_\ell\right)/d_\ell\right)\bigg] ,
    \label{5}
\end{equation}
where $\mathrm{erf}$ is the error function. The following parameter values were used: $\ell_\mathrm{min} = 0.01R_\odot$, $x_\ell = 0.735$ and $d_\ell = 0.02$. The dependencies of the original (\ref{4}) and corrected (\ref{5}) mixing-lengths on the radius are shown in Fig.\,\ref{f1}.

\begin{figure}
\centering
\includegraphics[width=\linewidth]{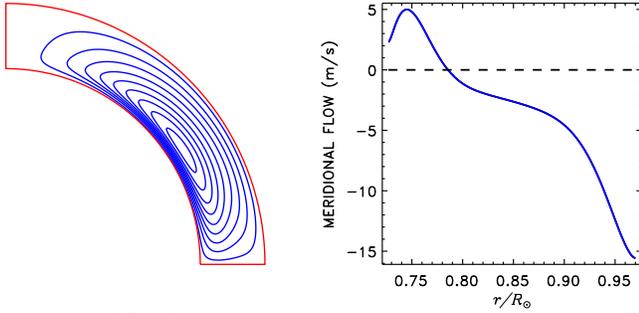}
\caption{Meridional flow in the differential rotation model with corrected
    $\ell$ of the Eq.\,(\ref{5}). {\sl Left:} stream lines of the flow. {\sl Right:} meridional velocity as the function of the radius at  $45^\circ$ latitude. A positive velocity means an equatorward flow.}
    \label{f2}
\end{figure}

Correction of the mixing length was motivated by difficulties in dynamo modeling. The velocity of the meridional flow is sensitive to the value of $\ell$ and decreases with a decrease in this value.
Meridional flow velocity at the bottom of the convection zone reaches 10 m/s when Eq.\,(\ref{4}) is used. The time of magnetic field advection by the meridional flow controls the cycle period in dynamo models. The computed cycles are always shorter than 11 years for near-bottom flow of 10 m/s. The meridional flow computed with the corrected mixing length (\ref{5}) is shown in Fig.\,\ref{f2}. The flow amplitude near the bottom boundary of about 5 m/s is acceptable for dynamo modeling. It may be noted that the flow in Fig.\,2 is closer to the latest helioseismological data of Rajaguru \& Antia (2015) than former results for not corrected $\ell$ (fig.\,2 of Kitchatinov \& Olemskoy 2011a).

Computed differential rotation is shown in Fig.\,3. The model computation agrees with the observed surface rotation and with seismological data. Our model, however, does not include the tachocline - the thin layer of transition from latitude-dependent to rigid rotation. Helioseismology finds that the tachocline thickness does not exceed 4\% of the solar radius and its central radius $r_\mathrm{c} = (0.693\pm 0.002)R_\odot$ (Charbonneau et al. 1999). The tachocline, therefore, lies below the inner radius of the convection zone of $r_\mathrm{i} = 0.713R_\odot$ (Chistensen-Dalsgaard et al. 1991; Basu \& Antia 1997). Moreover, only the {\sl radial} inhomogeneity of rotation, producing a toroidal field from the {\sl radial} field, is large in the tachocline. The radial field should be small at the base of the convection zone. Otherwise, it is a relic field penetrating from the radiative zone. Contrary to what is frequently stated, the tachocline cannot be important for dynamo.

\begin{figure}\centering
\includegraphics[width=\linewidth]{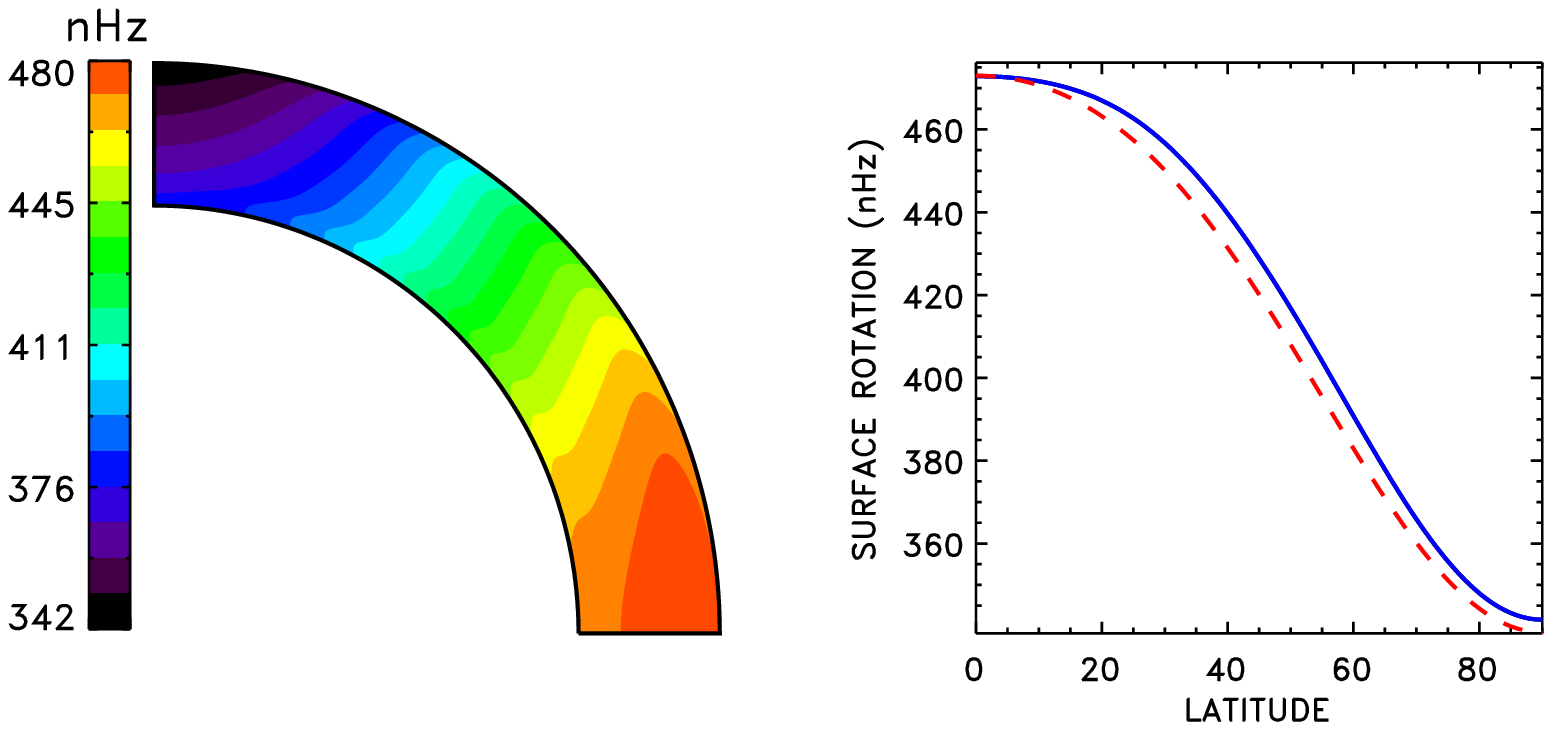}
\caption{The differential solar rotation of our model. \\ {\sl Left:} Angular
    velocity iso-contours. {\sl Right:} dependence of the surface rotation frequency on latitude. The dashed line shows the Doppler measurements by Snodgrass \& Ulrich (1990).}
    \label{f3}
\end{figure}

A special feature of the model considered is that the turbulent transport coefficients are not prescribed, as it is usually done, but are expressed in terms of the inhomogeneity of entropy, which in turn is governed by the Eq.\,(\ref{2}). The eddy viscosities, in particular, read
\begin{equation}
    \nu_n = \nu_{_\mathrm{T}}\phi_n(\Omega^*) ,\ \ \
    \nu_{_\mathrm{T}} = -\frac{\tau\ell^2 g}{15 c_\mathrm{p}}\frac{\partial S}{\partial r} ,
    \label{6}
\end{equation}
where $\nu_{_\mathrm{T}}$ is the isotropic viscosity for non-rotating fluid. The dependence on the rotation rate enters via the functions $\phi_n(\Omega^*)$ of the Coriolis number
\begin{equation}
    \Omega^* = 2\tau\Omega ,
    \label{7}
\end{equation}
where $\tau$ is the correlation time of turbulent convection. The viscosity in rotating fluid is anisotropic and it is defined by five viscosity coefficients (\ref{6}) for $n = 1,2,...,5$ (Kitchatinov et al. 1994). Such an approach reduces uncertainty in specifying the model parameters, especially in stellar applications. We keep following the same strategy in the dynamo model but specify the parameter $\mathrm{Pm}$ of proportionality between the turbulent viscosity and magnetic diffusivity,
\begin{equation}
    \nu_{_\mathrm{T}} = \mathrm{Pm}\ \eta_{_\mathrm{T}} ,\ \ \
    \eta_{_\mathrm{T}} = -\frac{\tau\ell^2 g}{\mathrm{Pm} 15 c_\mathrm{p}}\frac{\partial S}{\partial r} ,
    \label{8}
\end{equation}
by fitting observations. Numerical experiments by Yousef et al. (2003) show that the magnetic Prandtl number $\mathrm{Pm}$ is of order one but its precise value for convective turbulence is not certain.
\section{Dynamo model}
\subsection{Basic equations}
The large-scale magnetic field $\vec B$ in our model is governed by the mean-field induction equation (see, e.g., Krause \& R\"adler 1980)
\begin{equation}
    \frac{\partial{\vec B}}{\partial t} = {\vec\nabla}\times\left(
    {\vec V}\times{\vec B} + {\vec{\cal E}}\right) .
    \label{9}
\end{equation}
In this equation, $\vec V$ is the large-scale velocity (\ref{3}), ${\vec{\cal E}} = \langle{\vec u}\times{\vec b}\rangle$ is the mean electromotive force (EMF) due to the correlation of fluctuating velocity $\vec u$ and magnetic field $\vec b$, the angular brackets signify averaging over, e.g., longitude. For the simplest case of isotropic and homogeneous turbulence, the EMF  ${\vec{\cal E}} = \alpha{\vec B} - \eta_{_\mathrm{T}}{\vec\nabla}\times{\vec B}$ includes the (isotropic) turbulent diffusion and the $\alpha$-effect (Krause \& R\"adler 1980).

For stellar convective envelopes, turbulence inhomogeneity and rotation are essential. The inhomogeneity results in a \lq diamagnetic pumping'  of large-scale fields into the region of relatively low turbulence intensity near the base of the convection zone (Zel'dovich 1957; Krause \& R\"adler 1980). Allowance for this effect is important for solar dynamo modeling. Rotation induces anisotropy in the turbulence. Different eddy transport coefficients apply to the directions along and across the rotation axis, while the direction and magnitude of the diamagnetic transport velocity attains a dependence on the orientation of the magnetic field relative to the rotation axis. The expression for EMF of inhomogeneous and rotating fluids is rather complicated (cf., e.g., Pipin 2008). To simplify matters, we separate the contributions of diffusion (${\vec{\cal E}}^\mathrm{diff}$), diamagnetic pumping (${\vec{\cal E}}^\mathrm{dia}$), and alpha-effect (${\vec{\cal E}}^\alpha$) in the EMF:
\begin{equation}
    {\vec{\cal E}} = {\vec{\cal E}}^\mathrm{diff}
        + {\vec{\cal E}}^\mathrm{dia}
        + {\vec{\cal E}}^\alpha .
    \label{10}
\end{equation}

The diffusive part of the EMF reads
\begin{eqnarray}
    {\vec{\cal E}}^\mathrm{diff} &=& -\eta {\vec\nabla}\times{\vec B}
        - \eta_\| {\vec e}\times ({\vec e}\cdot{\vec\nabla}){\vec B} ,
    \nonumber \\
    &&\eta = \eta_{_\mathrm{T}}\phi(\Omega^*),\ \ \
    \eta_\| = \eta_{_\mathrm{T}}\phi_\|(\Omega^*),
    \label{11}
\end{eqnarray}
where $\eta$ is the isotropic part of the diffusivity, $\vec e$ is the unit vector along the rotation axis, $\eta_\|$ is an additional diffusivity along this axis, $\eta_{_\mathrm{T}}$ is the diffusivity (\ref{8}) for non-rotating fluid, and the dependence on rotation rate is included by  the functions $\phi$ and $\phi_\|$ of the Coriolis number (\ref{7}). The explicit expressions for these functions are given by Kitchatinov et al. (1994).

Diamagnetic transport in a rotating fluid is specified by the following equation
\begin{eqnarray}
    {\vec{\cal E}}^\mathrm{dia} &=& -({\vec\nabla}\tilde{\eta})\times{\vec B}
        + ({\vec\nabla}\eta_\|)\times{\vec e}({\vec e}\cdot{\vec B}) ,
    \nonumber \\
    &&\tilde{\eta} = \eta_{_\mathrm{T}}\phi_1(\Omega^*),
    \label{12}
\end{eqnarray}
where $\eta_\|$ is the same diffusivity as in Eq.\,(\ref{11}) and the explicit expression for the function $\phi_1(\Omega^*)$ was derived in our paper concerning diamagnetic pumping and its implications for the solar dynamo (Kitchatinov \& Nepomnyashchikh 2016).

Radius $r_\mathrm{i}$ of the base of the convection zone is defined with the condition for the radiative heat flux to fit stellar luminosity $L$: $F^\mathrm{rad} = L/(4\pi r_\mathrm{i}^2)$. This is also the inner radius of the domain of simulation for both the differential rotation and dynamo. This radius computed for the Sun, $r_\mathrm{i} = 0.727R_\odot$, is larger than the value of $r_\mathrm{i} = 0.713R_\odot$ detected by helioseismology (Christensen-Dalsgaard et al. 1991; Basu \& Antia 1997). The difference is because our model does not account for penetration of convection into the region of subadiabatic stratification. The penetration region of low diffusivity is, however, important for the dynamo (Kitchatinov \& Olemskoy 2012b). To take the region into account, we decreased the turbulent diffusivity in a thin layer near the inner boundary,
\begin{equation}
    \eta_{_\mathrm{T}} =  \frac{1}{\mathrm{Pm}}\left[ \nu_\mathrm{i} + \frac{1}{2} (\nu_{_\mathrm{T}} - \nu_\mathrm{i})
    \bigg( 1 +\mathrm{erf}\left(\left({r}/{r_\mathrm{i}} - x_\eta \right)/d_\eta
    \right)\bigg)\right]
    \label{13}
\end{equation}
similar to the decreased mixing length of Fig.\,1. In this equation, Pm is the magnetic Prandtl number (\ref{8}), $\nu_{_\mathrm{T}}$ is the turbulent viscosity of Eq.\,(\ref{6}), $\nu_\mathrm{i} = 10^{-4}\times\nu_{_\mathrm{T}}^\mathrm{max}$, where $\nu_{_\mathrm{T}}^\mathrm{max}$ is the largest viscosity value within the convection zone, $x_\eta = 1.1$ and $d_\eta = 0.025$. The dependencies of the diffusivities (\ref{11}) and (\ref{12}) of our model on radius are shown in Fig.\,4.

\begin{figure}
\includegraphics[width=8truecm]{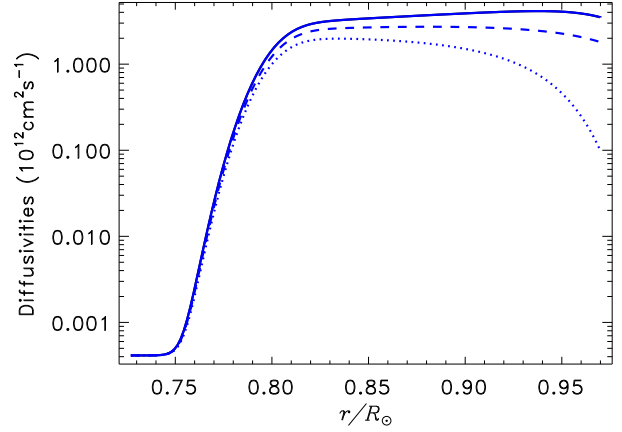}
\caption{The dependencies of the coefficients $\eta$ (full line),
    $\eta_\|$ (dotted) and $\tilde\eta$ (dashed) on the heliocentric radius for \\ Pm = 3.}
    \label{f4}
\end{figure}

The large-scale magnetic field $\vec B$ can be split in its toroidal and poloidal parts similar to Eq.\,(\ref{3}) for the large-scale velocity:
\begin{equation}
    {\vec B} = {\vec e}_\phi B
    + {\vec\nabla}\times\left({\vec e}_\phi\frac{A}{r\sin\theta}\right) .
    \label{14}
\end{equation}
Substitution of this equation to (\ref{9}) provides two equations for the toroidal and poloidal fields. The poloidal field equation,
\begin{equation}
    \frac{\partial A}{\partial t} = \frac{1}{\rho r^2\sin\theta}\left(
    \frac{\partial\psi}{\partial r}\frac{\partial A}{\partial\theta} -
    \frac{\partial\psi}{\partial\theta}\frac{\partial A}{\partial r}\right)
    + r\sin\theta\ {\cal E}_\phi ,
    \label{15}
\end{equation}
includes only the azimuthal component ${\cal E}_\phi$ of EMF. The contributions by turbulent diffusion and diamagnetic pumping to ${\cal E}_\phi$ can be found from  Eqs.\,(\ref{11}) and (\ref{12}). We do not give explicit expressions for these rather bulky contributions. The contribution by the $\alpha$-effect,
\begin{equation}
    {\cal E}^\alpha_\phi = \alpha \frac{B(r_\mathrm{i},\theta)}{1 + (B(r_\mathrm{i},\theta)/B_0)^2}
    F(\theta)\phi_\alpha(r/r_\mathrm{e}) ,
    \label{16}
\end{equation}
corresponds to its particular version known as the Babcock-Leighton mechanism. This mechanism is non-local in space (Choudhuri et al. 1995; Durney 1955): the poloidal field near the top of the convection zone is produced from the bottom toroidal field $B(r_\mathrm{i},\theta)$.

The function $\phi_\alpha$ defines the near-surface region of poloidal field generation. It is the same function as used formerly by Kitchatinov \& Olemskoy (2012b):
\begin{equation}
    \phi_\alpha(x) = \frac{1}{2}\left( 1 + \mathrm{erf}((x + 2.5h_\alpha -1)/h_\alpha)\right) .
    \label{17}
\end{equation}
$h_\alpha = 0.02$ in the computations to follow. The function
\begin{equation}
    F(\theta) = \cos\theta\sin^{n_\alpha}\theta
    \label{18}
\end{equation}
defines the $\alpha$-effect profile in latitude. The profile is anti-symmetric about the equator and $n_\alpha$ controls the degree of its equatorial concentration. In all probability, the Babcock-Leighton mechanism operates in the Sun (Erofeev 2004; Dasi-Espuig et al. 2010; Kitchatinov \& Olemskoy 2011b). However, the corresponding $\alpha$-effect has not been derived \lq from first principles' so that heuristic equations (\ref{16}) - (\ref{18}) have to be used. The $n_\alpha$ parameter will be specified by the model tuning.

The azimuthal component of the induction equation (\ref{9}) gives the equation for the toroidal field. With allowance for Eqs. (\ref{3}) and (\ref{14}), we find
\begin{eqnarray}
    \frac{\partial B}{\partial t} &=&
    \frac{1}{r^2\rho}\frac{\partial\psi}{\partial r}
        \frac{\partial}{\partial\theta}\left(\frac{B}{\sin\theta}\right)
    -\frac{1}{r\sin\theta}\frac{\partial\psi}{\partial\theta}
        \frac{\partial}{\partial r}\left(\frac{B}{\rho r}\right)
    \nonumber \\
    &+& \frac{1}{r}\left(\frac{\partial\Omega}{\partial r}\frac{\partial A}{\partial\theta}
    - \frac{\partial\Omega}{\partial\theta}\frac{\partial A}{\partial r}
    + \frac{\partial(r{\cal E}_\theta)}{\partial r}
    - \frac{\partial{\cal E}_r}{\partial\theta}\right) .
    \label{19}
\end{eqnarray}
Differential rotation generates the toroidal field more efficiently than the $\alpha$-effect.
The contributions by turbulent diffusion (\ref{11}) and diamagnetic pumping (\ref{12}) only are, therefore, kept in the components ${\cal E}_r$ and ${\cal E}_\theta$ of the EMF (\ref{10}) in Eq.\,(\ref{19}), the contribution by the $\alpha$-effect being neglected (the $\alpha\Omega$-approximation). Only the azimuthal component (\ref{16}) of the vector $\vec{\cal E}^\alpha$ enters the dynamo equations (\ref{15}) and (\ref{19}).

The boundary conditions correspond to an interface with a superconductor at the inner boundary,
\begin{equation}
    {\cal E}_\theta = 0,\ A = 0,\ r = r_\mathrm{i} ,
    \label{20}
\end{equation}
and to the \lq quasi-vacuum' condition (vertical field),
\begin{equation}
    B = 0,\ \frac{\partial A}{\partial r} = 0,\ \ r = r_\mathrm{e} ,
    \label{21}
\end{equation}
at the external boundary.

The initial-value problem was considered. The initial toroidal field was put zero and the potential $A$ for the poloidal field was prescribed as
\begin{eqnarray}
    A_0(r,\theta) &=& \frac{B_N}{4(1 - r_\mathrm{i}/r_\mathrm{e})^2}
    \left( r_\mathrm{i}^2 + 2r_\mathrm{e}(r-r_\mathrm{i}) - r^2\right)
    \nonumber \\
    &&\times\left( 1 - P + \cos\theta (1 + P)\right)\sin^2\theta ,
    \label{22}
\end{eqnarray}
where $B_N$ is the initial field at the northern pole and $P$ is the parity index in the range of $-1\leq P\leq 1$. Dynamo equations allow two types of equatorial symmetry: symmetric (quadrupolar) and ani-symmetric (dipolar) modes of magnetic field. The initial field (\ref{22}) belongs to the dipolar parity for $P = -1$ and to the quadrupolar parity for $P=1$. Other $P$-values mean a superposition of dipolar and quadrupolar modes. After a sufficiently long time, solutions of the dynamo equations \lq forget' the initial conditions. The results to follow belong to such asymptotic regimes.
\subsection{Numericals}
Numerical solution of dynamo equations has to be coordinated with the differential rotation model supplying the differential rotation (Fig.\,3) and the meridional flow (Fig.\,2) for the dynamo. The finite-difference numerical grid of $N_r$ points on the radius,
\begin{eqnarray}
    r_j &=& \frac{1}{2}\left( r_\mathrm{e} + r_\mathrm{i} -
    (r_\mathrm{e} - r_\mathrm{i})\cos\left(\pi\frac{j-3/2}{N_r - 2}\right)\right), \nonumber \\
    &&2\leq j \leq N_r - 1 ,
    \nonumber \\
    &&r_1 = r_\mathrm{i}, r_{N_r} = r_\mathrm{e},
    \label{23}
\end{eqnarray}
is, therefore, common for both models. This grid is a liner transformation of zeros of Chebyshev polynomials from the range of [-1,1] into [$r_\mathrm{i},r_\mathrm{e}$]. The denser grid near the boundaries is necessary for resolution of boundary layers in the differential rotation model and for resolution of the fine magnetic structure near the bottom of the convection zone.

Two versions of numerical dynamo code were developed. The differential rotation model uses the Legendre polynomial expansion for dependencies on latitude. The same spectral method was used in the first version of the dynamo model. This led to an equation system for the expansion coefficients dependent on time and radius. The equation system is solved by the Crank-Nicholson method of second-order accuracy in time and radius (Press et al. 1992). However, if the magnetic diffusivity near the base of the convection zone is so small, or meridional velocity is so large, that the magnetic Reynolds number $\mathrm{Rm} = V^\mathrm{m}r_\mathrm{i}/\eta_{_\mathrm{T}} \gsim 10^4$, this version of the model suffers from the Gibbs phenomenon (Press et al. 1992) and the series in Legendre polynomials converge very slowly. For these cases, the second version of the numerical model, which uses a finite-difference grid on the latitude as well, was applied. The grid of $N_\theta$ points is uniform in $\cos\theta$ but not uniform in latitude. The denser grid near the equator helps to resolve a finer magnetic structure of this region. To ensure numerical stability, the spatial derivatives in the radius are treated implicitly in the time-stepping of this second version of the model. The majority of the  results discussed below were obtained with this second version of the model. Computations with $N_r = 151$ and $N_\theta = 201$ provide sufficient resolution.

For $\mathrm{Rm} < 10^4$, the results of both codes were practically identical.
As a further test, the cases $A'$, $B'$ and $C'$ of the \lq dynamo benchmark'
by Jouve et al. (2008) were reproduced.
\newpage

\section{Results and discussion}
\subsection{Equatorial parity and cycle period}
Hydromagnetic dynamos can be understood as instabilities of the conducting fluid flows to seed magnetic fields (Krause \& R\"adler 1980). For an instability to develop, some controlling parameter should exceed a certain critical value (Chandrasekhar 1961). Such a controlling parameter in our model is the parameter $\alpha$ of equation (\ref{16}). Non-decaying magnetic fields emerge when this parameter exceeds its critical magnitude $\alpha_\mathrm{c}$. These threshold magnitudes differ between the equatorially symmetric ($\alpha_\mathrm{c}^\mathrm{q}$) and antisymmetric ($\alpha_\mathrm{c}^\mathrm{d}$) fields. The symmetry type with smaller $\alpha_\mathrm{c}$ usually prevails (this is always the case with our model). In other words, computations starting from the initial field (\ref{22}) of mixed parity eventually converge to pure parity corresponding to the smaller of the two values $\alpha_\mathrm{c}^\mathrm{d}$ or $\alpha_\mathrm{c}^\mathrm{q}$.

\begin{figure}
\includegraphics[width=8 truecm]{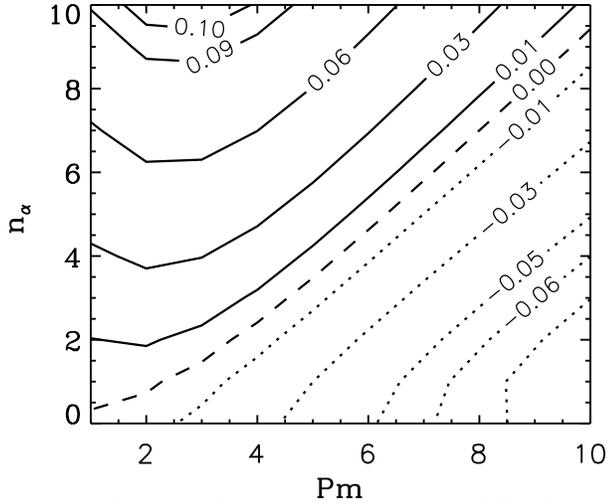}
\caption{Isolines of the symmetry index $s$ of Eq.\,(\ref{24}), controlling
    parity of the generated fields, on the plane of the parameters $\mathrm{Pm}$ and $n_\alpha$ of our model. The dashed line is the boundary between the region of positive $s$, where the solar-type equator-antisymmetric fields are dominant, and the region of quadrupolar fields.}
    \label{f5}
\end{figure}

Figure\,\ref{f5} shows isolines of the quantity
\begin{equation}
    s = \alpha_\mathrm{c}^\mathrm{q}/\alpha_\mathrm{c}^\mathrm{d} - 1
    \label{24}
\end{equation}
on the plane of the parameters $\mathrm{Pm}$ (\ref{8}) and $n_\alpha$ (\ref{18}) of our model.
Preference for the (antisymmetric) fields of dipolar parity is usually related to the relatively large diffusivity in the bulk of the convection zone (Chatterjee et al. 2004; Hotta \& Yokoyama 2010): quadrupolar modes of the poloidal field have a smaller latitudinal scale and are affected more strongly by dissipation. The predominance of dipolar modes for relatively small $\mathrm{Pm}$ in Fig.\,5 agree with this notion. This Figure also shows that the degree of equatorial concentration of the $\alpha$-effect is significant too. Dipolar modes dominate for relatively large $n_\alpha$ when $\alpha$-effect operates at low latitudes.

Large-scale solar fields are close to dipolar symmetry (St\-enflo 1988; Obridko et al. 2006). Their quadrupolar part comprises less then 10\% of magnetic flux and does not show the 11-year cycle (Stenflo 1988). The deviations from dipolar parity may be related to fluctuations in the $\alpha$-effect (Latyshev \& Olemskoy 2016).

\begin{figure}
\includegraphics[width=8 truecm]{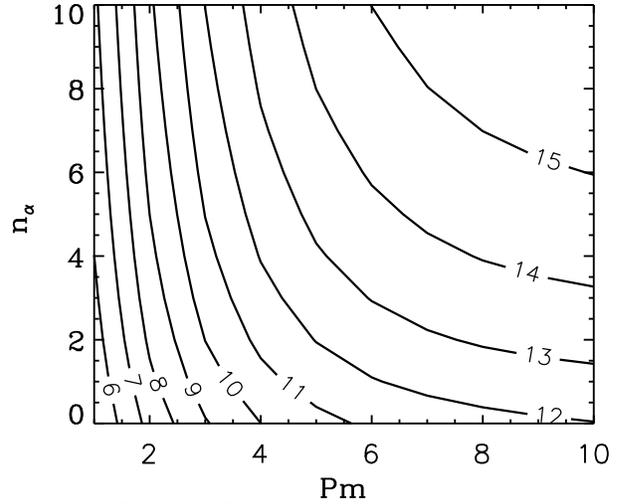}
\caption{Isolines of the activity cycle period on the same plane of
    parameters $\mathrm{Pm}$ and $n_\alpha$ as in Fig.\,5. The cycles duration in years is shown in the isoline gaps.}
    \label{f6}
\end{figure}

The region above the dashed line in Fig.\,5 corresponds to the equatorial symmetry of the solar magnetic fields. Further restrictions on the model parameters follow from a comparison with the cycle period. Figure 6 shows the periods computed for the \lq marginal' dipolar modes ($\alpha = \alpha_\mathrm{c}^\mathrm{d}$). The meridional circulation time is believed to control the magnetic cycle period. Figure 6, however, shows that there is a dependence on magnetic diffusivity also.

\subsection{The basic model}
Comparison of Figs. 5 and 6 shows that a rather narrow region of $2.5 < \mathrm{Pm} < 3.5$ and $n_\alpha > 2$ corresponds to both the 11-year cycle period and dipolar parity. The Babcock-Leighton type $\alpha$-effect is related to the solar active regions present at relatively low latitudes $\lambda < 30^\circ$ mainly. We therefore accept $n_\alpha = 7$ and $\mathrm{Pm} = 3$ for our \lq basic' model. The threshold value of $\alpha_\mathrm{c} = 0.152$ m/s corresponds to these parameters. Non-decaying magnetic fields are present for $\alpha$ exceeding this threshold. The amount of the excess can be inferred from stellar data.

\begin{figure*}[!t]\centering
\includegraphics[width=12 truecm]{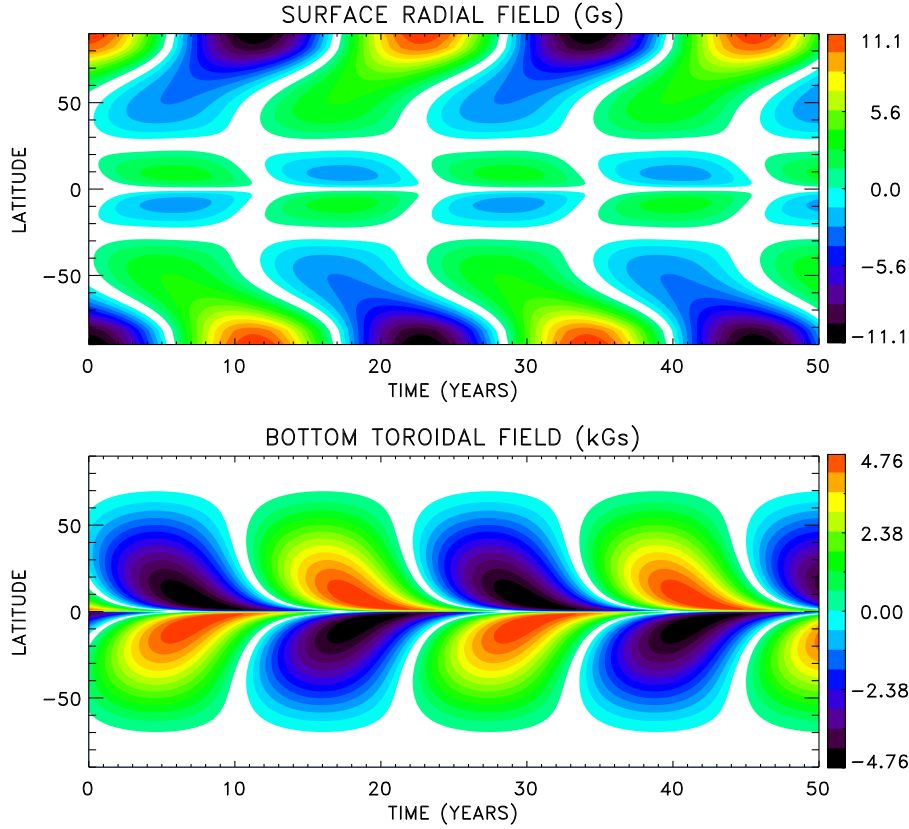}
\caption{Time-latitude diagrams of the surface radial field ({\sl top panel})
    and bottom toroidal field ({\sl bottom panel}) computed with the dynamo model.}
    \label{f7}
\end{figure*}

Generation of large-scale fields in convective envelopes of solar-type stars is related to the stars' rotation. The large-scale fields increase the effective radius of stellar wind emanation (Kraft 1967) thus leading to a decrease of the rotation rate with stellar age $t$ approximately by the Skumanich (1972) law $\Omega \propto t^{-1/2}$. The proportionality coefficient in this relation depends on temperature (mass) of a star. Specification of this dependence gave rise to gyrochronology -- determination of stellar ages from their rotation rates and effective temperatures (Barnes 2003). It was found recently, however, that main-sequence stars older than some billion years do not obey gyrochronology. They rotate faster than gyrochronology predicts (van\,Saders et al. 2016; Metcalfe et al. 2016). Such stars show rather low magnetic activity and practically stop spinning down. Rengarajan (1984) was probably the first to notice this spindown limit. Figure\,1 of Rengarajan paper clearly shows the maximum rotation period (dependent on $B-V$ colour of a star) where spindown stops. The rotation rate corresponding to this maximum period, by all probability, is marginal for the onset of large-scale dynamos. For the solar value of $B-V = 0.656$, fig.\,1 of Rengarajan (1984) gives the maximum rotation period of about 28.5 day. This is about 12\% above the current rotation period of 25.4 day. We, therefore, accept 12\% supercritical $\alpha = 0.17$~m/s also for the model of the solar dynamo. We also accept the value of $B_0 = 10^4$~Gs slightly above the equipartition field for the deep solar convection for the $B_0$ parameter of Eq.\,(\ref{16}).

\begin{figure}
\includegraphics[width=8 truecm]{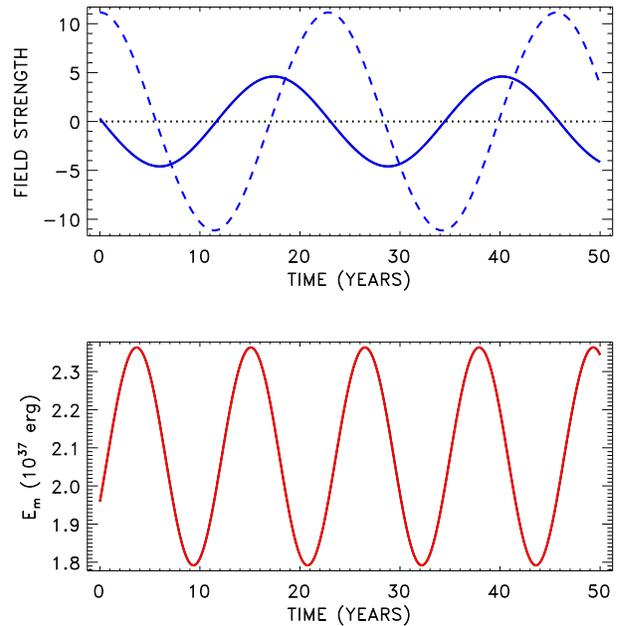}
\caption{{\sl Top panel:} the dashed line shows the radial field (in Gauss) at
    the northern pole for the same computation as Fig.\,7. The toroidal field strength (in kGs) at the base of the convection zone and 15$^\circ$ latitude is shown by the full line. {\sl Bottom:} total energy of toroidal field in the convection zone.}
    \label{f8}
\end{figure}

The time-latitude diagrams for the surface radial field and the bottom toroidal field for the basic model of the solar dynamo are shown in Fig.\,7. Magnitudes of these fields can be seen in Fig.\,\ref{f8} where the magnetic energy of the convection zone is also shown in line with the field strengths.

\begin{figure*}[!t]\centering
\includegraphics[width=12 truecm]{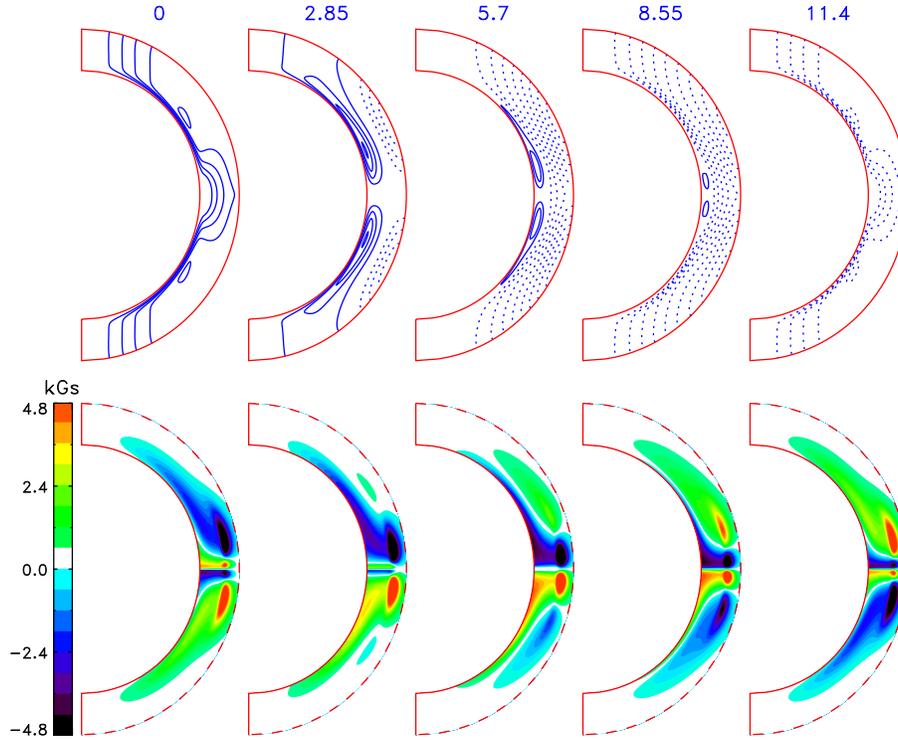}
\caption{{\sl Top panel:} poloidal field-lines for several instants of a
    magnetic cycle. Full and dashed lines correspond to the clockwise and anti-clockwise circulation, respectively. The numbers on the top show time in years in correspondence with Fig.7. {\sl Bottom panel:} toroidal field distributions. The upper (dashed) boundary shows the radius of $r = 0.81R_\odot$ below which the toroidal field is localised.}
    \label{f9}
\end{figure*}

The radial field attains its maximum strength at the poles and decreases rapidly with decreasing latitude. This polar concentration of the field, caused by convergence of the meridional flow to the poles (Fig.2), agrees with the magnetographic observations of Svalgaard et al. (1978). The toroidal field at the base of the convection zone amounts to several thousand Gauss. The combination of relatively weak radial fields at the surface with three orders of magnitude stronger toroidal fields at the bottom is brought about by the diamagnetic effect of turbulent convection. The diamagnetic pumping concentrates the meridional field to the base of the convection zone where it reaches a sufficient strength $\sim100$\,Gs for generation of strong toroidal fields. Otherwise, the differential rotation could not wind a thousand-Gauss toroidal field in the course of a solar cycle (Kitchatinov \& Nepomnyashchikh 2016).

The largest toroidal fields of Fig.7 are produced at low latitudes $\sim 10-15^\circ$, where the highest sunspot activity is observed (Vitinsky et al. 1986). Such a toroidal field distribution is caused by the equatorial meridional flow near the base of the convection zone (Hazra et al. 2014).

Reversals of the polar field in Figs 7 and 8 occur close to the instants of maximum strength of the near-base toroidal field but after a delay of about two years relative to the maxima of magnetic energy. The total magnetic energy in our model is about $10^{37}$~erg. This value is consistent with ob\-ser\-va\-ti\-on-based estimations by Cameron and Sh\"ussler (2015) who have shown that the total toroidal magnetic flux $F$, corresponding to the observed polar field of the Sun, reaches about $10^{24}$~Mx at the activity maxima. The corresponding magnetic energy can be estimated as $E_\mathrm{m} \sim RB_\phi F/4$, where $B_\phi$ is the characteristic strength of the toroidal field and $R$ is the characteristic radius of the field location. Taking $R = r_\mathrm{i} \simeq 5\times 10^{10}$~cm, $B_\phi \sim 10^3$~Gs and $F \sim 10^{24}$~Mx, we find $E_\mathrm{m} \sim 10^{37}$~erg in accord with Fig.8. It may be noted that reversal of the sign of the $\alpha$-effect (i.e., $\alpha = - 0.17$~m/s) in our model leads to a steady dynamo with magnetic energy about 50 times larger compared to its amplitude in the cyclic dynamo with positive $\alpha$. This circumstance is important for the theory of Grand maxima of solar/stellar activity (Kitchatinov \& Olemskoy 2016).

We conclude by noting that the tuning of two parameters with the condition of cor\-res\-pon\-den\-ce to the observed period of the solar cycle and equatorial symmetry of the large-scale fields brings the model into agreement with other observed characteristics of large-scale organisation of solar activity. The coordination of the dynamo and differential rotation models has prepared their application to solar-type stars.
\phantomsection
\section*{Acknowledgments}
This work was supported by the Russian Foundation
for Basic Research (project 16--02--00090).
\phantomsection
\section*{References}
\begin{description}
\item{} Barnes,~S.\,A. 2003, \apj\ {\bf 586}, 464
\item{} Barnes,~J.\,R., Collier\,Cameron,~A., Donati,~J.-F., James,~D.\,J., Marsden,~S.\,C., \& Petit,~P. 2005, \mnras\ {\bf 357}, L1
\item{} Basu,~S., \& Antia,~H.\,M. 1997, \mnras\ {\bf 287}, 189
\item{} Cameron,~R., \& Sch\"ussler,~M. 2015, \sci\ {\bf 347}, 1333
\item{} Chandrasekhar,~S. 1961, Hydrodynamic and Hydromagnetic Stability
    (Oxford: Clarendon Press)
\item{} Charbonneau,~P., Christensen-Dalsgaard,~J., Henning,~R., Larsen,~R.\,M., Schou,~J., Thompson,~M.\,J., \& Tomczyk,~S. 1999 \apj\ {\bf 527}, 445
\item{} Chatterjee,~P., Nandy,~D., \& Choudhuri,~A.\,R. 2004, \aap\ {\bf 427}, 1019
\item{} Choudhuri,~A.\,R., Sch\"ussler,~M., \& Dikpati,~M. 1995 \aap\ {\bf 303}, L29
\item{} Christensen-Dalsgaard,~J., Gough,~D.\,O., \& Thompson,~M.\,J.\ 1991 \apj\ {\bf 378}, 413
\item{} Croll,~B., Walker,~G.\,A.\,H., Kuschnig,~R., Matthews,~J.\,M., Rowe,~J.\,F., Wal\-ker,~A., Ru\-c\-in\-ski,~S.\,M., Hatzes,~A.\,P., et al. 2006, \apj\ {\bf 648}, 607
\item{} Dasi-Espuig,~M., Solanki,~S.\,K., Krivova,~N.\,A., Cameron,~R., \& Pe\~{n}uela,~T. 2010, \aap\ {\bf 518}, A7
\item{} Durney,~B.\,R. 1995, \sp\ {\bf 160}, 213
\item{} Durney,~B.\,R. 2003, \sp\ {\bf 217}, 1
\item{} Erofeev,~D.\,V. 2004, in: Multi-Wavelength Investigation of Solar Activity, IAU Symp. 223 (Eds. A.\,V.~Stepanov, E.\,E.~Benevolenskaya, A.\,G.~Kosovichev, Cambridge Univ. Press), p.97
\item{} Hazra,~G., Karak,~B.\,B., \& Choudhuri,~A.\,R. 2014, \apj\ {\bf 782}, 93
\item{} Hotta,~H., \& Yokoyama,~T. 2010, \apjl\ {\bf 714}, L308
\item{} Jouve,~L.\, Brun,~A.\,S., Arlt,~R., Brandenburg,~A., Dikpati,~M., Bonanno,~A., K\"apyl\"a,~P.\,J., Moss,~D., et al. 2008, \aap\ {\bf 483}, 949
\item{} Kitchatinov,~L.\,L., Pipin,~V.\,V., \& R\"udiger,~G. 1994, \an\ {\bf 315}, 157
\item{} Kitchatinov,~L.\,L. 2002, \paj\ {\bf 28}, 626
\item{} Kitchatinov,~L.\,L., \& Olemskoy,~S.\,V. 2011a, \mnras\ {\bf 411}, 1059
\item{} Kitchatinov,~L.\,L., \& Olemskoy,~S.\,V. 2011b, \paj\ {\bf 37}, 656
\item{} Kitchatinov,~L.\,L., \& Olemskoy,~S.\,V. 2012a, \mnras\ {\bf 423}, 3344
\newpage
\item{} Kitchatinov,~L.\,L., \& Olemskoy,~S.\,V. 2012b, \sp\ {\bf 276}, 3
\item{} Kitchatinov,~L.\,L., \& Nepomnyashchikh,~A.\,A. 2016, \\ \adsr\ {\bf 58}, 1554
\item{} Kitchatinov,~L.\,L.\, \& Olemskoy,~S.\,V. 2016, \mnras\ {\bf 459}, 4353
\item{} Kitchatinov,~L.\,L. 2016, \ga\ {\bf 56} (in press); arXiv: 1603.07852
\item{} Kraft,~R.\,P. 1967, \apj\ {\bf 150}, 551
\item{} Krause,~F., \& R\"adler,~K.-H.
    1980, Mean-Field Magnetohydrodynamics and Dynamo Theory (Oxford: Pergamon Press)
\item{} Latyshev,~S.\,V., \& Olemskoy,~S.\,V. 2016, \paj\ {\bf 42}, 488
\item{} Metcalfe,~T.\,S., \& Egeland,~R., \& van\,Saders,~J. 2016, \apjl\ {\bf 826}, L2
\item{} Obridko,~V.\,N., Sokoloff,~D.\,D., Kuzanyan,~K.\,M., Shelting,~B.\,D., \& Za\-kh\-a\-rov,~V.\,G. 2006, \mnras\ {\bf 365}, 827
\item{} Pipin,~V.\,V. 2008, \gafd\ {\bf 102}, 21
\item{} Press,~W.\,H., Teukolsky,~S.\,A., Vetterling,~W.\,T., \& Flannery,~B.\,P. 1992, Numerical Recipies (Cambridge Univ. Press)
\item{} Rajaguru,~S.\,P., \& Antia,~H.\,M. 2015, \apj\ {\bf 813}, 114
\item{} Rengarajan,~T.\,N. 1984, \apjl\ {\bf 283}, L63
\item{} R\"udiger,~G. 1989, Differential Rotation and Stellar Convection (New York: Gordon \& Breach)
\item{} van\,Saders,~J.\,L., Ceillier,~T., Metcalfe,~T.\,S., Silva\,Aguirre,~V., Pinsonneault,~M.\,H., Garc\'{i}a,~R.\,A., Mathur,~S., \& Davies,~G.\,R. 2016, \nat\ {\bf 529}, 181
\item{} Schou,~J., Antia,~H.\,M., Basu,~S., Bogart,~R.\,S., Bush,~R.\,I., Chitre,~S.\,M., Chri\-sten\-sen-Dalsgaard,~J., Di\,Mauro,~M.\,P., et al. 1998, \apj\ {\bf 505}, 390
\item{} Skumanich,~A. 1972, \apj\ {\bf 171}, 565
\item{} Snodgrass,~H.\,B., \& Ulrich,~R.\,K. 1990, \apj\ {\bf 351}, 309
\item{} Stenflo,~J.\,O. 1988, \apss\ {\bf 144}, 321
\item{} Svalgaard,~L., Duvall,~T.\,L., \& Scherrer,~P.\,H. 1978, \sp\ {\bf 58}, 225
\item{} Vitinsky,~Yu.\,I., Kopecky,~M., \& Kuklin,~G.\,V. 1986, The Statistics of Sunspots (Moscow: Nauka) [in Russian]
\item{} Walker,~G.\,A.\,H., Croll,~B., Kuschnig,~R., Walker,~A., Rucinski,~S.\,M., Mat\-thews,~J.\,M., Guenther,~D.\,B., Moffat,~A.\,F.\,J., et al. 2007, \apj\ {\bf 659}, 1611
\item{} Wilson,~P.\,R., Burtonclay,~D., \& Li,~Y. 1997, \apj\ {\bf 489}, 395
\item{} Yousef,~T.\,A., Brandenburg,~A., \& R\"udiger,~G. 2003, \aap\ {\bf 411}, 321
\item{} Zel'dovich,~Ya.\,B. 1957, Sov. Phys. JETP {\bf 4}, 460
\end{description}
\end{document}